\documentclass[epj]{webofc}
\usepackage[utf8]{inputenc}
\usepackage[varg]{txfonts}   
\usepackage{booktabs}
\usepackage{xcolor}
\definecolor{darkred}{rgb}{0.4,0.0,0.0}
\definecolor{darkgreen}{rgb}{0.0,0.4,0.0}
\definecolor{darkblue}{rgb}{0.0,0.0,0.4}
\usepackage[bookmarks,linktocpage,colorlinks,
    linkcolor = darkred,
    urlcolor  = darkblue,
    citecolor = darkgreen]{hyperref}
\usepackage{subfigure}
\wocname{EPJ Web of Conferences}
\woctitle{Lattice2017}
\newcommand{\be}{\begin{equation}}
\newcommand{\ee}{\end{equation}}

\begin{document}
%
\selectlanguage{english}
\title{Instanton-dyon ensembles
reproduce deconfinement and
chiral restoration phase
transitions
}
\author{
\firstname{Edward} \lastname{Shuryak}\thanks{The author emphasizes that this talk is based on works done in
collaboration with R.Larsen, I.Zahed and Y.Liu.}
}
\institute{%
Department of Physics and Astronomy, Stony Brook University, \\Stony Brook NY 11794 USA
}
\abstract{%
 Paradigm shift in gauge topology at finite temperatures, from the instantons to their constituents 
--  instanton-dyons -- has recently lead to studies of their ensembles and very significant advances. 
Like instantons, they have fermionic zero modes, and their collectivization
at sufficiently high density  explains the {\em chiral symmetry breaking transition}.
Unlike instantons, these objects have  electric and magnetic charges. Simulations of 
the instanton-dyon ensembles have demonstrated that their back reaction on the Polyakov line  
modifies its potential and generates the {\em deconfinement phase transition}. 
For the $N_c=2$ gauge theory the transition is second order, for QCD-like theory with
 $N_c=2$ and two light quark flavors $N_f=2$  both transitions
are weak crossovers at happening at about the same condition. Introduction of quark-flavor-dependent
periodicity phases (imaginary chemical potentials)  leads to drastic changes in both
transitions. In particulaly, in the so called $Z(N_c)-QCD$ model
the deconfinement transforms to strong first order transition,  while the chiral condensate 
does not disappear at all. The talk will also cover more detailed studies of correlations
between the dyons, effective eta' mass and other screening masses.  
}
\maketitle
\section{Introduction}
Since I   participate in Lattice conference series
intermittedly, I try to put in these few pages a very wide range of issues. 
Section \ref{topo} emphasizes  topology-induced phenomena, such as chiral symmetry breaking, and technical-but-crucially-important limitations they pose for lattice simulations. 
Section \ref{dyons}  introduces the instanton-dyons and section \ref{ensembles} analytical and numerical studies of their ensembles. Those discussions are rather brief, for
a bit more detailed discussion see 2016 proceedings \cite{1610.08789} and original papers.
  Calculation of hadronic correlators
 using instanton-dyon ensembles, and the lesson followed, are summarized  in section \ref{correlators}.
 
 Perhaps it is worth to remind lattice community about the ``Gauge topology" workshop series. 
 The first took place in August 2015 at Symons center, Stony Brook, the second in 
 Nov.2016 at ECT*,Trento. The third will happen 
 May 28-June 01 2018, also at  ECT*,Trento: the interested readers take a note.
\section{Why is understanding of the gauge topology so important for lattice community?}\label{topo}
One general reason is that
 the topological solitons, once observed and studied on the lattice, can then be used
to understand many other important phenomena. Some of them cannot be done in Euclidean time
 settings, requiring {\em real time} kinetics. 
 E.g. the monopoles were long studies on the lattice, with a confirmation (see e.g. \cite{D'Alessandro:2010xg}) that
  they do Bose condensed below $T_c$. The monopoles   not only explain confinement, but also were used to explain small viscosity (mean free path) \cite{Ratti:2008jz}  and strong 
jet quenching \cite{Xu:2014tda,Ramamurti:2017zjn}   in quark-gluon plasma, especially near the transition temperature $T_c$.

Another reason is that understanding properties of  topological objects on the lattice is in fact nacessary
to answer many lattice questions, 
among them such basic ones  as: ``How large  should be the lattices  used in practice?"

In order to explain this point, let me briefly remind the history of chiral symmetry breaking.
In a classic paper Nambu and Jona-Lasinio showed that an attraction in the scalar $\bar q q$ channel,
if strong enough, can dynamically ``gap" the surface of the Dirac sea. The origin
of this interaction -- claimed to be the origin of the mass of the ``constinuent quarks",
the nucleons and thus ourselves -- was of course in 1961 completely unknown.
In my paper \cite{Shuryak:1981ff},  it was suggested that
NJL attraction is nothing else but 't Hooft effective $2N_f$ interaction
induced by instantons. Two parameters of the NJL model has been substituted by another two: the
total instanton-antiinstanton density $n\equiv1/R^4\sim 1/fm^4$ and their typical size $\rho \sim 1/3 \, fm$.  
Of course, 't Hooft vertex does more than the NJL operator:  in particular, it breaks the $U(1)_A$ 
symmetry.

Statistical mechanics of instanton ensemble,  including  't Hooft interaction
 to all orders,  known as  
 the Interacting Instanton Liquid Model,  has been developed and solved numerically in 1990's, for a review
  see
\cite{Schafer:1996wv}.  Among other things, it introduced the notion of ``collectivized zero mode zone", or ZMZ for short. The width of ZMZ is of the order of the typical matrix element
of a ``quark hopping matrix" made of overlaps of zero modes
of neighboring instantons. Its magnitude is of the order of
\be \Delta \lambda_{ZMZ} \sim {\rho^2 \over R^3}  \sim 20 \, MeV
\ee
(for $\rho,R$ of the instanton ensemble mentioned above). 

While these features and values of the parameters have been confirmed on the lattice many times, lattice community still had not appreciated its importance, and keeps rediscovering it.
For example, Graz group had observed \cite{Glozman:2012hw}, that (i) most fluctuations in simulations come
from few Dirac eigenmodes with eigenvalues $ | \lambda |< \Delta \lambda_{ZMZ}$; (ii) their removal
from propagators lead to significant $\sim 30-50\%$ changes of hadronic masses; (iii) in a way that both the $SU(N_f)$ and $U(1)a$ chiral symmetries get restored. 

Several plenary talks at this (and certainly earlier) lattice conferences discussed the requirements for the lattice sizes, assuming the smallest scale appearing in quark propagators
is $m_\pi/2$. This would indeed be true when quark masses go to zero, but 
{\em not in the real world!}. Indeed, 
$m_\pi/2\approx 70\, MeV$,  significantly $exceeding$ the ZMZ width $ \Delta \lambda_{ZMZ}$ (which
b.t.w. is essentially independent on the box size and quark masses).  Multiple failures of chiral extrapolation fits to various quantities over the years attest to the same statement. It is known and published for at least two  decades, yet the lesson
has not yet been learned. 

 Another aspect at this issue can be explained as follows. The number of
ZMZ  eigenmodes  is nothing but the mean number of instantons on the lattice,
their density times the lattice volume. With current
lattice spacings $a\sim .1 \, fm$, the
lattice volumes of the order of $(2\, fm)^4$ 
may look sufficient:
 and yet they include on average only about $16=2^4$ instantons,
or one instanton and one antiinstanton per direction. Not surprising, one finds large
fluctuations and severe finite volume effects. Instanton liquid simulations were done with larger boxes, typically with $500$  instantons. Sorry to say, (much more expensive)
 lattice simulations need such sizes of the box as well. 

In conclusion, the gauge topology provides a serious additional argument in favor of the proposal,
suggested by Luscher at this meeting, to use lattices {\em as large as possible},
even if the available resources would only be enough for a single or few configurations. 

\section{Instanton-dyons}\label{dyons}
The so called Polyakov line is used as a deconfinement order parameter, being nonzero at $T>T_c$.
Interpreting this as existence of nonzero average $A_0$ field, one needs to modify all clasical solutions
respectively.  When such solutions were found in 1998 \cite{Kraan:1998sn,Lee:1998bb} it has been realized
that 
instantons  get split into $N_c$ (number of colors) constituents, the selfdual {\em instanton-dyons}\footnote{
They are  called  ``instanton-monopoles" in applications to supersymmetric settings, e.g. 
by Khose et al and Unsal et al. Similar (but no idenical) objects were called 
 the ``instanton quarks" by Zhitnitsky et al.} , 
 connected only by (invisible) Dirac strings.
Since these objects have nonzero electric and magnetic charges and source
Abelian (diagonal) massless gluons, the corresponding ensemble is 
an ``instanton-dyon plasma", with long-range Coulomb-like forces between constituents.  

The first application of the instanton-dyons were made soon after their discovery
in the context of supersymmetric gluodynamics \cite{Davies:1999uw}. This paper solved a puzzling
mismatch of the value of the gluino condensate, between the instanton-based and general supersymmetric
evaluations of it. 

Diakonov and collaborators (for review see \cite{Diakonov:2009jq} )
 emphasized that, unlike the (topologically protected) instantons, the dyons interact directly with
 the holonomy field. They suggested that since such dyon (anti-dyon)  become denser
at low temperature, their back reaction  may overcome perturbative holonomy potential and drive it
to its confining value, leading to  vanishing of the mean Polyakov line, or confinement.
Specifically, Diakonov and collaborators focused on the self-dual sector $L,M$ and studied the one-loop
contribution to the partition function \cite{Diakonov:2004jn}. The volume element of the moduli space was
written in terms of dyons coordinates as a determinant of certain matrix $G$, to be referred to as Diakonov determinant. In a dilute limit it leads to
 Coulomb interactions between the dyons, but in the dense region it becomes strongly repulsive, till at certain density
the moduli volume vanishes. 

  A semi-classical  confining regime has been defined by Poppitz et al
~\cite{Poppitz:2011wy,Poppitz:2012sw}   in a carefully devised setting of softly broken supersymmetric models.
 While the setting includes a compactification on a small circle, with  weak coupling and
 an {\em exponentially  small}  density of dyons, the minimum at the confining holonomy
  value is induced by the repulsive interaction in the dyon-antidyon molecules (called  
 $bions$ by these authors). 
 The crucial role of the supersymmetry is the cancellation of the perturbative Gross-Pisarski-Yaffe-Weiss (GPYW)  \cite{Gross:1980br} holonomy potential:
 as a result, in this setting there is no deconfined phase with trivial holonomy at all, unless supersymmetry is softly broken.
  Sulejmanpasic and myself \cite{Shuryak:2013tka} proposed a simple analytic model for the dyon ensemble
with  dyon-antidyon ``repulsive cores", and have shown how they may naturally
 induce confinement in dense enough dyonic ensemble.

Recent progress to be discussed below is related to studies of the instanton-dyon ensembles.
We will focus on a series of papers devoted to high-density phase and mean field
approximation \cite{Liu:2015ufa,Liu:2015jsa,Liu:2016thw,Liu:2016mrk,Liu:2016yij} 
and on the direct
numerical simulation of the dyon ensembles \cite{Faccioli:2013ja,Larsen:2014yya,Larsen:2015vaa,Larsen:2015tso,Larsen:2016fvs} in section \ref{ensembles}. 

Periodicity condition along the Matsubara circle can be defined with some arbitrary angles $\psi_f$ for quarks
with the flavor $f$. 
As was determined by van Baal and collaborators, fermionic zero mode ``hops" from one type of dyon to the next
at certain critical values. The resulting rule is: it belogs to the dyon corresponding to the segment of the
holonomy circle $\nu_i$ to which the periodicity phase belongs: $\mu_i<\psi_f<\mu_{i+1}$.

In physical QCD all quarks are fermions, so $\psi_f=\pi$ for all $f$. Therefore all $N_f$ quarks  have zero modes bound to one and the same dyon type, the $L$-dyons.
But one can introduce other arrangements of these phases. In particular, for $N_c=N_f$, the opposite extreme is 
the so called  {\em $Z(N_c)$ QCD}, proposed in \cite{Kouno:2012zz}. These authors
suggested to imply imaginary chemical potentials in such a way, as to put
the fermionic periodicity phases  symmetrically around the circle.
In the theory of the  instanton-dyons this means that 
 each  dyon type has one -- its ``own" -- quark flavor which binds to it.

\section{Instanton-dyon ensembles} \label{ensembles}

The first direct simulation of the instanton-dyon ensemble with dynamical fermions
has been made by Faccioli and myself in \cite{Faccioli:2013ja}. The general setting
follows the example of the ``instanton liquid", it included the determinant'of the so called "hopping matrix", a part of the Dirac operator in the quasizero-mode
sector. It has been done for $SU(2)$ color group and the number of fermions
flavors $N_f=1,2,4$. Except in the last case, the chiral symmetry breaking phase transitions
have been clearly observed, for dense enough dyon ensemble.

 Larsen and myself  \cite{Larsen:2015vaa} use direct numerical simulation of the instanton-dyon ensemble, both in the high-T dilute and low-T dense regime. 
 Unlike the previous work, it uses
 classical dyon-antidyon interaction determined in Ref.\cite{Larsen:2014yya}.
 The holonomy potential as a function
of all parameters of the model is determined and minimized.
At high density of the dyons their back reaction shifts the minimum to $\nu=1/2$,
which is the confining value for SU(2) ($cos(\pi \nu)=0$):
the confinement transition is thus generated .
 The self-consistent parameters of the
ensemble, minimizing the free energy, is  determined for each density. 
The transition is second order, as expected based on lattice results.

The next work of Larsen and myself \cite{Larsen:2015tso} addressed the issue of chiral symmetry breaking 
in the $N_c=2$ theory with two light quark flavors $N_f=2$.
The results for the mean Polyakov line are shown in Fig.1 (left) by blue filled circles
as a function of the parameter $S$, the ``instanton action", proportional to the log of the  temperature
\be S=({11N_c \over 3}- {2N_f \over 3})log({T \over \Lambda})
\ee
So, larger $S$ at the r.h.s. of the figure correpond to high $T$, and thus to more dilute dyon ensemble,
since densities contain $exp(-S_i)$.
The measure behavior of Polyakov line VEV in this theory indicates a smooth cross over transition.

 This numerical simulations are done for dynamical quarks, with a partition function appended by the fermionic determinant
evaluated in the zero mode approximation. 
Using two sizes of the system, with 64 and 128 dyons, we 
identify the finite-size effects in the eigenvalue distribution, and 
extrapolate to infinite size system.
The 
location of the chiral transition temperature is 
defined both by extrapolation of the quark condensate, from below, and
the so called ``gaps" in the Dirac spectra, from above. The values of the 
quark condensate are shown by black triangles in the right side of Fig.1.

We do indeed observe,  for this QCD-like SU(2) gauge theory with 2 flavors of light fundamental quarks, that  the deconfinement 
and chiral symmetry restoration transitions occur about at the same 
 dyon density,   namely $S=7.5-8$ (see short vertical lines shown for convenience).
  Determination of the precise transition points  
 is difficult 
  since  both transitions appear to be in this case just a smooth crossovers.
  By the way, the dyon actions are in 
 the confined phase a half of it, so the action per dyon  in the simulations is about $4\hbar$.
 This four is the large parameter of the semiclassical approximation used.

The $Z(N_c)$ QCD has been studied  in the mean field framework \cite{Liu:2016yij},  
by statistical simulations \cite{Larsen:2016fvs} and also by lattice simulations \cite{Misumi:2015hfa}.
The first two papers consider the $N_c=N_f=2$ version of the theory, while the last one focus on the
$N_c=N_f=3$. In the former case the set of phases are $\psi_f=0,\pi$, so one quark is a boson and one is a fermion. In the latter  $\psi_f=\pi/3,\pi,-\pi/3$.

\begin{figure}[h]
\centering
\includegraphics[width=7.cm]{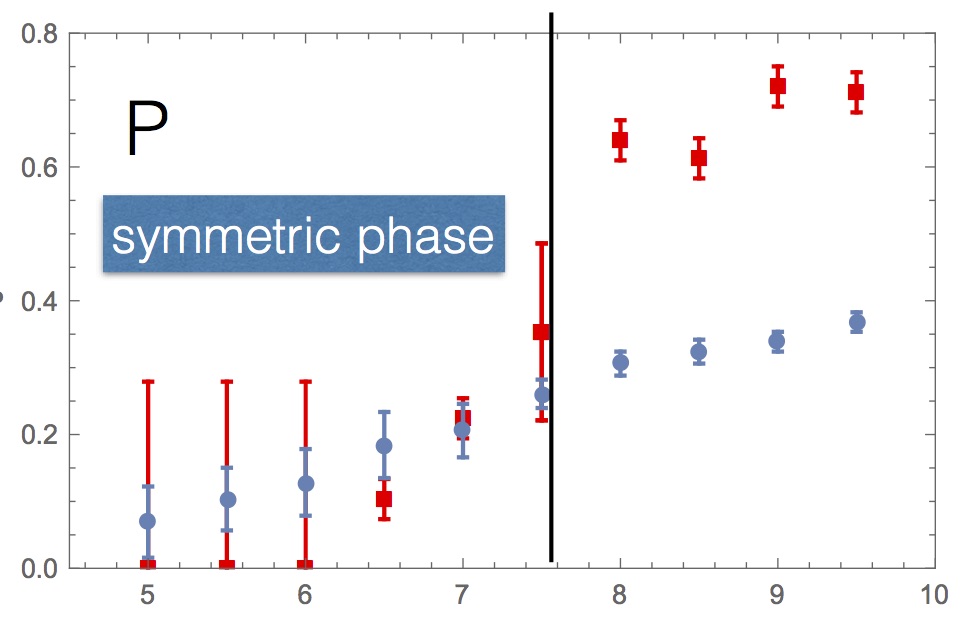}
\includegraphics[width=7.cm]{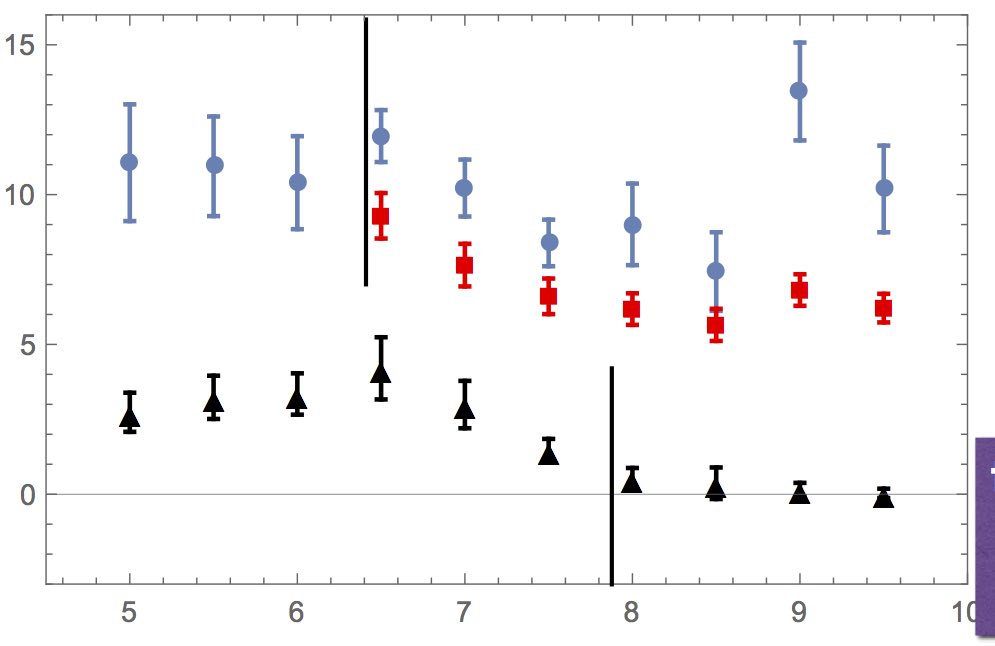}
\caption{(left) The mean Polyakov line P versus the density parameter $S$. Red squares
are for $Z_2 QCD$ while blue circles are for the usual QCD, both with $N_c=N_f=2$ .
(right) The quark condensate versus the density parameter $S$.
Black triangles correspond to the usual QCD: and they display chiral symmetry restoration.
Blue and red poins are for two flavor condensates of the $Z_2 QCD$: to the left of vertical
line there is a ``symmetric phase" in which both types of dyons and condensates are the same.
}
\label{fig_Z2}
\end{figure}

All  these works find that the deconfinement transition in $Z(N_c)$ QCD strengthens significantly, as compared to the usual QCD 
with the same $N_c,N_f$. In place of a smooth crossover, in \cite{Liu:2016yij}
find the second order transition, while our simulations \cite{Larsen:2016fvs} -- the red squares at the left side of Fig.\ref{fig_Z2} --
and those on the lattice \cite{Misumi:2015hfa} both see strong jumps in the Polyakov line VEV, indicated strong first order transition.

All three studies see a non-zero chiral condensates in the studied region of densities:
{\em none of them find any indications for a  chiral symmetry restoration}.  The values of the condensates for two quark flavors from our work  \cite{Larsen:2016fvs} are shown  by red squares and blue circles in 
the right side of  Fig.\ref{fig_Z2}. We also found that  
the spectrum of the Dirac eigenvalues has a very specific ``triangular" shape, characteristic
of a single-flavor QCD. This explains why the $Z(N_c)$ QCD has much larger condensate
than ordinary QCD, at the same dyon density, and also why there is no tendedecy to restoration.
As expected, all three works see that above the deconfinement transition there are
no symmetry between flavors, so they have 
different  chiral condensates, $<\bar u u>\neq <\bar d d>$. Also interesting that their
difference is markedly smaller than one could expect from the difference in the dyon densities.

\section{Hadronic  point-to-point correlation functions and instanton-dyons}  \label{correlators}
Let us start this last section with a brief discussion of a more general question:
{\em At which temperatures the description of gauge topology in terms of  the instanton-dyons
is expected to work? }

It is clear that that they should not be used at high enough $T$, because their whole
existence is based on presence of non-zero VEV of the Polyakov line, which is
absent at, say $T>\, 400 \, MeV$. And indeed, as we now know from recent lattice studies of the topological susceptibility and other considerations, at such high $T$ the gauge topology
is well represented by a {\em dilute instanton gas}.

The instanton-dyon description is most useful near the deconfinement-chiral phase transition
temperature, and this is what the discussion in the previous section was about.

At {\em sufficiently low} $T$ one, unfortunately, again has to abandon the instanton-dyon theory.
One reason for that is that the Polyakov line VEV and the dyon fields  both
have scales $O(T)$, and as $T\rightarrow 0$ those become small compared to
quantum scales related to $\Lambda_{QCD}$ and thus unimportant. 
Another way to say it is as follows: as the size of the Matsubara time box $\beta=1/T$ grows,
 the 3-dimensional objects --
 instanton-dyons -- must strongly interfere
  with each other, in order to cancel each-other's fields in such a manner as to, eventually, reproducing the instantons, which are
 4-dimensionally symmetric objects. These instantons also should have finite sizes and density as  $T\rightarrow 0$. In other words, the $T\rightarrow 0$
  limit is very singular for the instanton-dyon teory.

Now we  formulate the main question
of our
 paper \cite{1705.04707}: {\em Can the instanton-dyon theory, taken at its lowest end of temperatures $T\sim 100\, MeV$, reproduce known
phenomenology of hadronic correlation functions? }

The answer was affirmative, but with a nontrivial twist. It turned out, that it can only happen
if the instanton-dyons of $M_1M_2$ types are strongly correlated with the $L$-type dyons,
the only which have fermionic zero modes. These correlations are necessary to  
make these zero modes sufficiently well localized, and therefore strong enough,
to reproduce splittings between the channels of the correct magnitude. 

\begin{figure}[h]
\centering
\includegraphics[width=7.cm]{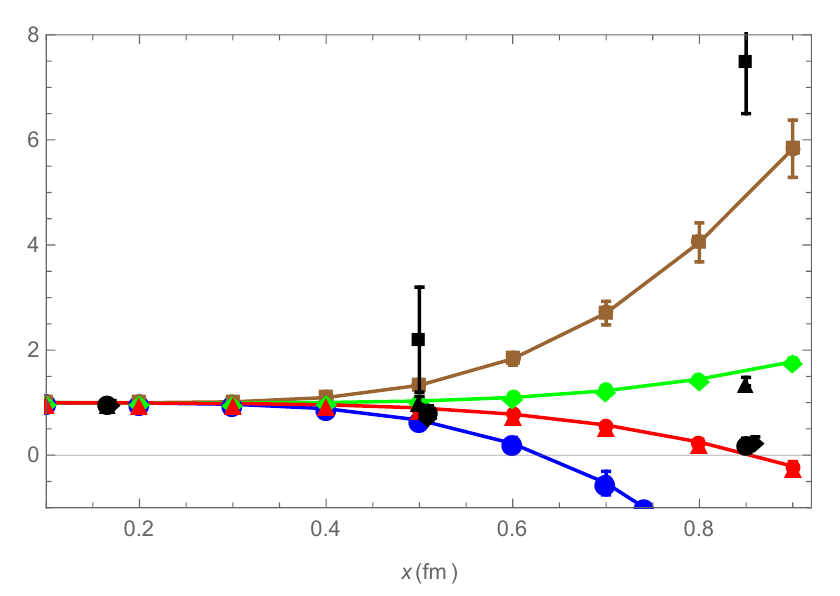}
\includegraphics[width=7.cm]{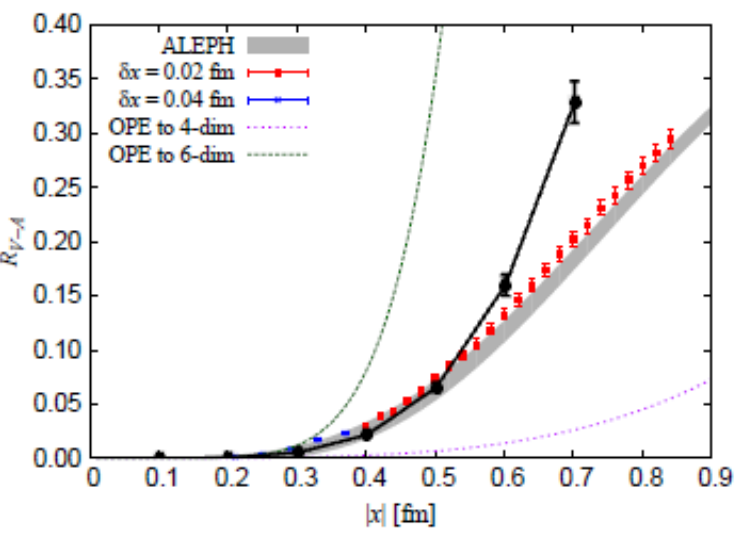}
\caption{(Left figure): 
The colored points are our results for the normalized point-to-point correlators, for four channels:  Pseudoscalar (Brown), Vector (Green), Axial vector (Red), and Scalar (Blue), top to bottom. The individual (black) points without lines are lattice data from \cite{Chu:1992mn}, their symbols are the same as for our data.
(Right figure) The normalized  vector-minus-axial 
 difference $(V (x) ? A(x))$  versus the distance x (fm). The narrow shadowed region corresponds to ALEPH data, the red and blue dots correspond to the lattice data \cite{Tomii:2017cbt}, for two lattice spacings indicated on the plot.
Our results are shown by black points.
}
\label{fig_C}
\end{figure}

These splittings are shown in Fig.\ref{fig_C}. The left figure shows point-to-point correlators
for four channels, normalized to those for free massless quarks (and thus all converging to 1
at small distances).  Note rather symmetric splittings between ``attractive" channels $PS,V$
when the non-perturbative effect is positive, and ``repulsive" channels $A,S$. Note also that
the $PS-S$ spliting, indicating violation of $U(1)_a$ chiral symmetry, is several times stronger than the $V-A$ splitting, indicating the violation of $SU(N_f)$ chiral symmetry.
This fact, known phenomenologically for a long time, is one of the strongest indication
of the important role of topology in hadronic spectroscopy.

These nice results, being in agreement with phenomenology and previous lattice studies, were
not observed for a generic random instanton-dyon ensemble. 
Indeed, it only happens if the $M-L$ dyon correlations are sufficiently strong.
The right figure indicates how we made the normalization of those 
 in our ensemble: the so called $ML$ dyon correlation parameter has been tuned to one particular combination, the $V-A$ (vector-minus-axial) correlator.
 While being weaker than the $PS-S$ splitting, as we already mentioned,  this one is rather accurately known, both from experiment
(the shaded area, corresponding to ALEPH data on the spectral densities) and 
also from recent lattice work \cite{Tomii:2017cbt}) (small red and blue points).

\clearpage
\bibliography{topo2}

\begin{thebibliography}{32}

\bibitem{1610.08789}
E.~Shuryak, EPJ Web Conf. \textbf{137}, 01018 (2017), \texttt{1610.08789}

\bibitem{D'Alessandro:2010xg}
A.~D'Alessandro, M.~D'Elia, E.V. Shuryak, Phys. Rev. \textbf{D81}, 094501
  (2010), \texttt{1002.4161}

\bibitem{Ratti:2008jz}
C.~Ratti, E.~Shuryak, Phys. Rev. \textbf{D80}, 034004 (2009),
  \texttt{0811.4174}

\bibitem{Xu:2014tda}
J.~Xu, J.~Liao, M.~Gyulassy, Chin. Phys. Lett. \textbf{32}, 092501 (2015),
  \texttt{1411.3673}

\bibitem{Ramamurti:2017zjn}
A.~Ramamurti, E.~Shuryak (2017), \texttt{1708.04254}

\bibitem{Shuryak:1981ff}
E.V. Shuryak, Nucl. Phys. \textbf{B203}, 93 (1982)

\bibitem{Schafer:1996wv}
T.~Schäfer, E.V. Shuryak, Rev. Mod. Phys. \textbf{70}, 323 (1998),
  \texttt{hep-ph/9610451}

\bibitem{Glozman:2012hw}
L.{\relax Ya}. Glozman, C.B. Lang, M.~Schrock, Acta Phys. Polon. Supp.
  \textbf{5}, 1001 (2012), \texttt{1207.7323}

\bibitem{Kraan:1998sn}
T.C. Kraan, P.~van Baal, Phys. Lett. \textbf{B435}, 389 (1998),
  \texttt{hep-th/9806034}

\bibitem{Lee:1998bb}
K.M. Lee, C.h. Lu, Phys. Rev. \textbf{D58}, 025011 (1998),
  \texttt{hep-th/9802108}

\bibitem{Davies:1999uw}
N.M. Davies, T.J. Hollowood, V.V. Khoze, M.P. Mattis, Nucl. Phys.
  \textbf{B559}, 123 (1999), \texttt{hep-th/9905015}

\bibitem{Diakonov:2009jq}
D.~Diakonov, Nucl. Phys. Proc. Suppl. \textbf{195}, 5 (2009),
  \texttt{0906.2456}

\bibitem{Diakonov:2004jn}
D.~Diakonov, N.~Gromov, V.~Petrov, S.~Slizovskiy, Phys. Rev. \textbf{D70},
  036003 (2004), \texttt{hep-th/0404042}

\bibitem{Poppitz:2011wy}
E.~Poppitz, M.~Unsal, JHEP \textbf{07}, 082 (2011), \texttt{1105.3969}

\bibitem{Poppitz:2012sw}
E.~Poppitz, T.~Schäfer, M.~Unsal, JHEP \textbf{10}, 115 (2012),
  \texttt{1205.0290}

\bibitem{Gross:1980br}
D.J. Gross, R.D. Pisarski, L.G. Yaffe, Rev. Mod. Phys. \textbf{53}, 43 (1981)

\bibitem{Shuryak:2013tka}
E.~Shuryak, T.~Sulejmanpasic, Phys. Lett. \textbf{B726}, 257 (2013),
  \texttt{1305.0796}

\bibitem{Liu:2015ufa}
Y.~Liu, E.~Shuryak, I.~Zahed, Phys. Rev. \textbf{D92}, 085006 (2015),
  \texttt{1503.03058}

\bibitem{Liu:2015jsa}
Y.~Liu, E.~Shuryak, I.~Zahed, Phys. Rev. \textbf{D92}, 085007 (2015),
  \texttt{1503.09148}

\bibitem{Liu:2016thw}
Y.~Liu, E.~Shuryak, I.~Zahed, Phys. Rev. \textbf{D94}, 105011 (2016),
  \texttt{1606.07009}

\bibitem{Liu:2016mrk}
Y.~Liu, E.~Shuryak, I.~Zahed, Phys. Rev. \textbf{D94}, 105012 (2016),
  \texttt{1605.07584}

\bibitem{Liu:2016yij}
Y.~Liu, E.~Shuryak, I.~Zahed, Phys. Rev. \textbf{D94}, 105013 (2016),
  \texttt{1606.02996}

\bibitem{Faccioli:2013ja}
P.~Faccioli, E.~Shuryak, Phys. Rev. \textbf{D87}, 074009 (2013),
  \texttt{1301.2523}

\bibitem{Larsen:2014yya}
R.~Larsen, E.~Shuryak, Nucl. Phys. \textbf{A950}, 110 (2016),
  \texttt{1408.6563}

\bibitem{Larsen:2015vaa}
R.~Larsen, E.~Shuryak, Phys. Rev. \textbf{D92}, 094022 (2015),
  \texttt{1504.03341}

\bibitem{Larsen:2015tso}
R.~Larsen, E.~Shuryak, Phys. Rev. \textbf{D93}, 054029 (2016),
  \texttt{1511.02237}

\bibitem{Larsen:2016fvs}
R.~Larsen, E.~Shuryak, Phys. Rev. \textbf{D94}, 094009 (2016),
  \texttt{1605.07474}

\bibitem{Kouno:2012zz}
H.~Kouno, Y.~Sakai, T.~Makiyama, K.~Tokunaga, T.~Sasaki, M.~Yahiro, J. Phys.
  \textbf{G39}, 085010 (2012)

\bibitem{Misumi:2015hfa}
T.~Misumi, T.~Iritani, E.~Itou, PoS \textbf{LATTICE2015}, 152 (2016),
  \texttt{1510.07227}

\bibitem{1705.04707}
R.~Larsen, E.~Shuryak, Phys. Rev. \textbf{D96}, 034508 (2017),
  \texttt{1705.04707}

\bibitem{Chu:1992mn}
M.C. Chu, J.M. Grandy, S.~Huang, J.W. Negele, Phys. Rev. Lett. \textbf{70}, 255
  (1993), \texttt{hep-lat/9211019}

\bibitem{Tomii:2017cbt}
M.~Tomii, G.~Cossu, B.~Fahy, H.~Fukaya, S.~Hashimoto, T.~Kaneko, J.~Noaki
  (JLQCD), Phys. Rev. \textbf{D96}, 054511 (2017), \texttt{1703.06249}

\end{thebibliography}

\end{document}